\documentclass[11pt]{elsart}

\usepackage{epsfig}
\usepackage{color}
\usepackage{amssymb}
\usepackage{graphicx}

\newcommand{\mcal}[1]{\mathcal{#1}}
\newcommand{\mcalE}{\eta}

\def\eg{e.g.$\!$}
\def\ee{{\rm e}}
\def\ii{{\rm i}}
\renewcommand{\vec}[1] {{\mathbf{#1}}}
\newcommand{\vt} {\vartheta}
\newcommand{\bea} {\begin{eqnarray}}
\newcommand{\eea} {\end{eqnarray}}
\newcommand{\beann} {\begin{eqnarray*}}
\newcommand{\eeann} {\end{eqnarray*}}
\newcommand{\labs} {\left\vert}
\newcommand{\rabs} {\right\vert}
\newcommand{\lsb} {\left[}
\newcommand{\rsb} {\right]}
\newcommand{\lrb} {\left(}
\newcommand{\rrb} {\right)}
\newcommand{\lcb} {\left\{}
\newcommand{\rcb} {\right\}}
\newcommand{\lab} {\left\langle}
\newcommand{\rab} {\right\rangle}

\newcommand{\ds} {\displaystyle}

\begin{document}
\begin{frontmatter}
\journal{Chemical Physics (invited)}

\title{Fingerprints of mesoscopic leads in the conductance of a molecular wire}
\author[MPIPKS]{Gianaurelio Cuniberti\corauthref{cor1}}, 
\corauth[cor1]{Corresponding author. Tel.: +49 (0)351 871 2213; fax: +49 (0)351 871 1999.}
\ead{cunibert@mpipks-dresden.mpg.de}
\author[MPIPKS,UNIR]{Giorgos Fagas}, and
\author[UNIR]{Klaus Richter}
\date{30 October 2001}
\address[MPIPKS]{Max Planck Institute for the Physics of Complex Systems,\\ N{\"o}thnitzer Str. 38, D-01187 Dresden, Germany}
\address[UNIR]{Institute for Theoretical Physics, University of Regensburg, 
\\ D-93053 Regensburg, Germany%
}

\begin{abstract}
The influence of contacts on linear transport
through a molecular wire attached to mesoscopic
tubule leads is studied. It is shown that low
dimensional leads, such as carbon nanotubes, in
contrast to bulky electrodes, strongly affect
transport properties. By focusing on the specificity
of the lead-wire contact, we show, in a fully
analytical treatment, that the geometry of this
hybrid system supports a mechanism of channel
selection and a sum rule, which is a distinctive
hallmark of the mesoscopic nature of the electrodes.
\end{abstract}
\end{frontmatter}

\maketitle

\section{Introduction}

Future electronic miniaturization may enter a regime where devices are
dominated by quantum mechanical laws and eventually reach the single-molecule
scale~\cite{Taur97}. 
Although molecular materials for electronics have already been
realized~\cite{CDLFBLSKL00+SKBB00+Ziemelis98+Tour96}, their implementation in
real applications~\cite{EL00,TKS98,G-GMLOL97} still has to cope with challenges in 
utilization, synthesis, and assembly~\cite{Landauer96c}.
Concerning theoretical ideas and methods the problem is also two-sided: On the
one hand many theoretical ideas are already footed on past pioneering work, such as the 
proposal of molecular rectification in 1974~\cite{AR74a} which was experimentally 
realized only 20 years later~\cite{WB93+MSA93}. On the other hand most of the 
conventional methods frequently employed for characterizing transport properties in 
microelectronic devices, such as the Boltzmann equation,
can no longer be applied at the molecular scale. Here conductance properties have 
to be calculated by employing full quantum mechanical approaches and by 
including the electronic structure of the molecules involved.

Electron transport on the atomic and molecular scale became a topic of intense
investigation since the invention of the scanning tunneling microscope (STM).
More recently, studies of transmission properties of single molecular junctions 
contacted to metallic leads~\cite{RZMBT97,PBdeVD00}
have intensified the interest in the basic mechanisms of conduction across molecular bridges.
The archetype of such a molecular device can be viewed as a 
donor and acceptor lead coupled by a molecule acting as a bridge. 
In such systems the traditional picture of electron transfer between donor and acceptor 
species is re-read in terms of a novel view in which a molecule can bear an electric 
current~\cite{Nitzan01a}.
Molecular bridges have been realized out of single organic
molecules~\cite{RZMBT97,ROBWML01}, short DNA strands~\cite{PBdeVD00}, but also as 
atomic wires~\cite{%
vanRuitenbeek01,%
YRGvdBAvR98,%
OKT98%
}.
Generally, contact effects alter the ``intrinsic conductance´´ of the molecules
in such experiments and call for closer theoretical studies.

In a parallel development the use of carbon nanotube (CNT) conductors has been the 
focus of intense experimental and theoretical activity as another promising direction 
for building blocks of molecular-scale circuits~\cite{KKKCSRO01,RKJTCL00+YWTA01}.
Carbon nanotubes exhibit a wealth of properties depending on their 
diameter, orientation of graphene roll up, and on their topology, namely
whether they consist of a single cylindrical surface~(single-wall) or many
surfaces~(multi-wall)~\cite{SDD98+McEuen00}.
If carbon nanotubes are attached to other materials to build elements of
molecular circuits, the characterization of 
contacts~\cite{TMDSRPNB01,ADX00+dePGWADR99} 
becomes again a fundamental issue. 
This problem arises also when a carbon nanotube is attached to another 
{\em molecular wire}, a single molecule or a molecular cluster with a privileged direction 
of the current flow.

In the usual theoretical treatment of transport through molecular wires, 
the attached leads are approximated by a continuum of free or quasi-free states, 
mimicking the presence of large reservoirs provided by bulky electrodes. 
However such an assumption may become inadequate when considering
leads with lateral dimensions of the order of the bridged molecule, as
for CNTs~\cite{TMDSRPNB01}.
The latter have been recently used as wiring elements~\cite{RKJTCL00+YWTA01}, as active 
devices~\cite{RKJTCL00+YWTA01,DMAA01+PTYGD01+MSSHA98}, and, attached to scanning
tunneling microscope (STM) tips, for enhancing their resolution~\cite{WMSS01,WJWCL98}. 
With a similar arrangement the fine structure of a twinned DNA molecule has been 
observed~\cite{NKANHYT00}. However, CNT-STM images seem to strongly depend on the 
tip shape and nature of contact with the imaging 
substrate~\cite{ANKN01+OBKMRGvdDR00+VdePHMLYMMRF98}. This calls for a better 
characterization of the contact chemistry of such hybrid structures.

This paper addresses the influence of the molecular wire-electrode contacts
on the linear conductance when the spectral structure and the geometry
of the electrodes plays an important r{\^o}le.
This allows us to quantify to which extent {\it mesoscopic leads} may affect
the conductance. Owing to the relevance of CNT-based devices, we focus on bridges 
between tubular leads. In previous density-functional-theory based treatments
the conductance through systems such as a C$_{60}$ molecule in between 
two CNTs has been calculated with high accuracy~\cite{CGFGRS01}. These numerical
approaches showed a strong sensitivity of the current on the system geometries and
strength of the molecule-CNT couplings. 

\begin{figure}[t]
\phantom{.}\vspace*{0.1cm}
\centerline{\epsfig{file=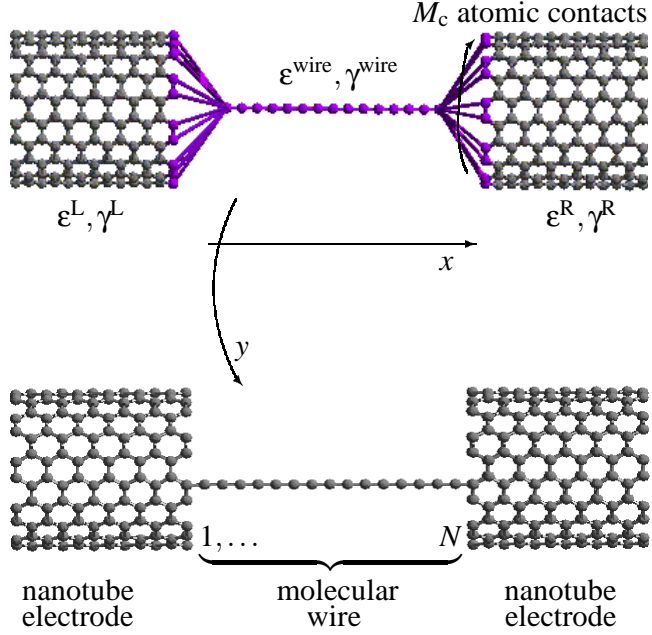, width=.60\linewidth}}
\phantom{.}\vspace*{1cm}
\caption{\label{fig:sketch} Schematic representation of the molecular 
wire-carbon nanotube hybrid with single (bottom) and multiple (top) contacts. 
on-site energies $\varepsilon^{\alpha={\rm L,R,wire}}$ are chosen to be zero.
}
\end{figure}

In the present paper we focus on such contact effects. As a simple model for
an atomic or molecular bridge we use a homogenous linear chain which enables 
us to derive analytical expressions for the conductance in a non-interacting electron approximation.
In addition, we implicitly assume that no significant charge is
transferred between the leads and the molecular bridge at equilibrium since
this could lead to an electrostatic potential-induced
inhomogeneity~\cite{LA00}. The latter may hold for an all-carbon~\cite{CGFGRS01} structure and
makes it possible 
to investigate the properties of our model in the whole parameter space.
The system exhibits distinct transport features depending on the number and 
strength of contacts between the molecular bridge and the interface as well as on 
the symmetry of the channel wave functions transverse to the transport direction. 
Our findings, which are common for leads with tube topology, are then studied in 
detail for CNT leads (Fig.~\ref{fig:sketch}.) by analytically treating the 
single-particle Green function.
In particular, we demonstrate on the one hand that configurations with only one 
molecule-lead contact activated give rise to complex conductance spectra 
exhibiting quantum features of both the molecule and the electrodes; 
on the other hand multiple contacts provide a mechanism for transport channel 
selection leading to a regularization of the conductance, entirely provided 
by topological arguments. 
Channel selection particularly highlights the r{\^o}le of molecular resonant 
states by suppressing details assigned to the electrodes.

\section{System and Method}

In a tight-binding description, the hamiltonian of the entire system,
$H = H_\mathrm{tubes} + H_{\rm wire} + H_{\rm coupling}$,
reads
\begin{eqnarray}
\label{eq:hamiltonian}
H &=& 
\sum_{\alpha={\rm L,R,wire}} \; 
\sum_{n^{\phantom{\prime}}_{\alpha},n_{\alpha}^\prime} 
t^\alpha_{n^{\phantom{\prime}}_{\alpha},n_{\alpha}^\prime}
{c_{n_\alpha^{\phantom{\prime}}}^{\alpha \dagger}}
c_{n_\alpha^\prime}^{{\alpha \phantom{ \dagger}}}
\\ && 
- \sum_{ m_{\rm L} \le M_{\rm c}} \Gamma_{m_{\rm L}}
\lrb
{{c_{m_{\rm L}}^{{\rm L} \dagger}}} c_1^{{\rm wire \phantom{ \dagger}}}
+ {\rm h.c.} \rrb
\nonumber
- \sum_{m_{\rm R} \le M_{\rm c}} \Gamma_{m_{\rm R}} 
\lrb
{c_{m_{\rm R}}^{{\rm R} \dagger}} c_N^{{\rm wire \phantom{ \dagger}}}
+ {\rm h.c.} \rrb \; .
\nonumber
\end{eqnarray}
Here, the matrix element $t^\alpha_{n^{\phantom{\prime}}_{\alpha},n_{\alpha}^\prime} = 
\varepsilon^{\alpha}_{n_{{\alpha}}} \delta_{n^{\phantom{\prime}}_{{\alpha}},n_{{\alpha}}^\prime} - \gamma^\alpha_{\left\langle n^{\phantom{\prime}}_{\alpha},n_{\alpha}^\prime \right\rangle}$ contains the on-site energy of each of the $n_ {\rm wire}=1,\dots, N$ chain-atoms, $\varepsilon^{\rm wire}$, the orbital energy relative to that of the lead atoms, $\varepsilon^{\rm L,R}$, and $\gamma^{\rm L,R}$, $\gamma^{\rm wire}$, and $\Gamma$ are nearest neighbour hopping terms between atoms of the left or right leads, molecular bridge, and the bridge/lead interface, respectively.
Note that $n_{\rm L,R}$ is a two-dimensional coordinate spanning the tube lattice.
Summations over $m_{\rm L}$ and $m_{\rm R}$ run over interfacial end-atoms of the leads.
In general, 
there are $M$ such atomic positions, defining the perimeter of the tube ends.
The number of hybridization contacts between a tube and the bridge
range between $M_{\rm c}=1$ (single contact case, SC) and $M_{\rm c}=M$ (multi-contact case,
MC).
Typical real molecular wires are $\pi$-conjugate carbon chains with thyol end
groups which in the present treatment are replaced by the linear chain model.

Since the major results we present are not qualitatively affected by the use of more realistic 
quantum chemical models which take into account the precise structure and properties of 
the molecular bridge and of the attached leads, we keep the description of our problem 
at the level of the tight-binding model.
In order to highlight the topological properties of tubular leads, we first study 
the simplest case in which 
periodic boundary conditions are imposed on semi-infinite square lattice stripes, 
with the cuts parallel to the lattice bonds. 
We call this electrode specie square lattice tubes (SLT).
In the case of CNT, when the graphene honeycomb lattice is rolled along the
lattice bonds such that $\ell$ hexagons are transversally wrapped, an 
armchair single wall $(\ell,\ell)$ nanotube is obtained, and $M=2\ell$.
According to Eq.~(\ref{eq:hamiltonian}) CNT are then described at the single-band tight-binding level for $\pi$ orbitals.
This is equivalent to assume that among the four valence carbon orbitals no interaction 
between the $\sigma (2s ~{\rm and}~2p_{x,y})$ and $\pi$ orbital is significant because 
of their different symmetries. The fourth electron, a $p_z$ orbital, determines the 
electronic properties which can be calculated by means of a tight-binding treatment,
on the same level as we treat the molecular bridge.
There are two such electrons per unit cell in a honeycomb structure, the $\pi$ and 
$\pi^*$ band, rendering the electronic properties of the material interesting, i.e.\
it can be {a priori} either metallic or semiconducting~\cite{SFDD92}.

We study quantum transport in the framework of the Landauer theory~\cite{IL99} 
which relates the conductance of the system in the linear response regime to an 
independent-electron scattering problem~\cite{FG99}. The electron wavefunction is 
assumed to extend coherently across the whole device.
The two-terminal conductance $g$ at zero temperature is simply proportional to the 
total transmittance, $T(E_{\rm F})$, for injected electrons at the Fermi energy 
$E_{\rm F}$: 
\begin{equation}
\label{eq:linear-zero-temperature-conductance}
g = \lrb {2 e^2}/ h \rrb T(E_{\rm F}) \; .
\end{equation}
The factor two accounts for spin degeneracy. The transmission function can be calculated 
from the knowledge of the molecular energy levels, the nature and the geometry of the contacts. 
One can see this by expressing the Green function matrix of the full
problem, 
$\vec{G^{\phantom{\rm wir}}}^{\!\!\!\!\!\!\!\! -1} = \vec{G^{\rm wire}}^{-1} + \vec{\Sigma^{\rm L} + \Sigma^{\rm R}}$,
in terms of the bare wire Green function and the self-energy
correction due to the presence of the leads. Making use of the Fisher-Lee
relation~\cite{FL81} one can finally write 
\begin{equation}
T (E) = 4 \; {\rm Tr} \lcb \vec{\Delta}^{\rm L} (E) {\vec{G}} (E) \vec{\Delta}^{\rm R} (E) \vec{G}^\dagger (E) \rcb ,
\end{equation}
where 
\begin{equation}
\vec{\Delta}^{\alpha} (E) = \frac \ii 2 \left.{\lrb {\vec{\Sigma}}^\alpha (z) 
-{{\vec{\Sigma}}^\alpha}^\dagger (z) \rrb } \rabs_{z=E + \ii 0^+} \, .
\end{equation}
For the system under investigation where only the first and last atom of the chain is 
coupled to the leads, the formula for the transmission simplifies to
\bea
\label{eq:transm-mit-spectral-densities}
T (E) =4 \ \Delta^{\rm L} (E) \Delta^{\rm R} (E) \labs G_{1N} \lrb E \rrb
\rabs^2 ,
\eea
where the spectral densities
$\Delta^{\rm L}$ and $\Delta^{\rm R}$ are the only non-zero elements 
$\lrb \vec{\Delta}^{\rm L}\rrb_{11}$ and $\lrb \vec{\Delta}^{\rm R}\rrb_{NN}$, respectively, 
of the matrices $\vec{\Delta}$.
The matrix element $\Delta^{{\rm L (R)}}$ is the spectral density of the
left (right) lead. It is related to the semi-infinite lead Green function 
matrix ${\mathcal{G}}^{\rm L(R)}$. It is minus the imaginary part of the 
lead self-energies (per spin),
\bea
\label{eq:spectr-density}
\Sigma^{\alpha}
=\Lambda^\alpha - \ii \; \Delta^\alpha
= \sum_{m_{\alpha}, m_{\alpha}^\prime} 
\Gamma^{\phantom{*}}_{m^{\phantom{\prime}}_{\alpha}} 
\Gamma^*_{m^\prime_{\alpha}} 
{\mathcal{G}}^\alpha_{ m_{\alpha}^{\phantom{\prime}} m_{\alpha}^\prime } \, ,
\eea
with $\alpha={\rm L,R}$. Owing to the causality of self-energy, its real part $\Lambda$ 
can be entirely derived from the knowledge of $\Delta$ via a Hilbert transform.

The rhs of Eq.~(\ref{eq:transm-mit-spectral-densities}) coincides with formulas
used to describe electron transfer in molecular systems~\cite{MKR94a+MKR94b}. 
The above relationship between the Landauer scattering matrix formalism on the one side
and transfer hamiltonian approaches on the other side has been worked out in the
recent past~\cite{Nitzan01a,HRHS00} showing {\it de facto} their equivalence. 
This enables us to make use of the formulas from a Bardeen-type picture in terms of 
spectral densities, which is often convenient for an understanding and
analysis of results obtained.

\section{Molecular Green function}
One has to calculate from the $N \times N$ matrix,
${G^{\rm wire}}^{-1} 
= E + \ii 0^+ - H_{\rm wire}$,
the Green function matrix element $G_{1N}$ needed in Eq.~(\ref{eq:transm-mit-spectral-densities}).
This matrix element refers to the two ends of the $N$-atom-molecule. Its computation 
requires an $N \times N$ matrix inversion.
Since only the molecular-end on-site energies are perturbed by the interaction with the 
leads via the self-energies $\Sigma^{\alpha}$, some general conclusions can already be 
drawn without an explicit computation of $G_{1N}$, namely one can write~\cite{%
SO93}
\bea
\label{eq:dressed-G1N}
G_{1 N} = \frac {G^{\rm wire}_{1 N}}{
\lrb 1 - \Sigma^{\rm L} G^{\rm wire}_{1 1} \rrb
\lrb 1 - \Sigma^{\rm R} G^{\rm wire}_{N N} \rrb
- 
\Sigma^{\rm L} \Sigma^{\rm R}
\lrb G^{\rm wire}_{1 N} \rrb^2
}.
\eea
The interaction with the leads dresses, via the self-energy
$\Sigma^{\alpha}$, the bare molecular wire Green function
element $G^{\rm wire}_{1 N}$. The latter 
can be calculated analytically in the case of a homogeneous wire 
($\varepsilon^{\rm wire}_n=\varepsilon^{\rm wire}$, 
$\gamma^{\rm wire}_{\left\langle 
n^{\phantom{\prime}}_{\alpha},n_{\alpha}^\prime \right\rangle}=\gamma^{\rm wire}$).
In fact, upon projecting on the $N$ dimensional molecular wire basis,
the determinant of the bare molecular Green matrix
factorizes into a dimensionless function of only the number of chain atoms,
and of the ratio $\mcalE^{\rm wire} = (E-\varepsilon^{\rm wire} )/(2 \gamma^{\rm wire})$. 
This leads to a closed form for the molecular contribution to the conductance. 
Namely, one can easily check that 
$G^{\rm wire}_{1N}=
{\gamma^{\rm wire}}^{N-1} {\rm det} \lrb G^{\rm wire} \rrb =
{(\gamma^{\rm wire})}^{-1} \xi_0/\xi_N$,
and
$G^{\rm wire}_{11} =G^{\rm wire}_{NN} ={\gamma^{\rm wire}}^{-1} \xi_{N-1} / \xi_N$,
where the exact form of $\xi$ reads:
\beann
\xi_N \lrb \mcalE^{\rm wire} \rrb =
{\lrb \mcalE^{\rm wire} + \sqrt{{(\mcalE^{\rm wire})}^2 -1} \rrb^{N + 1} - \lrb \mcalE^{\rm wire} - \sqrt{{(\mcalE^{\rm wire})}^2 -1} \rrb^{N + 1}}.
\eeann
After some algebra one finds that $\xi$ possesses the recursive property
$\xi_N {\xi_{N-2}} = { \xi^2_{N-1} -\xi^2_0}$,
which leads us to re-write Eq.~(\ref{eq:dressed-G1N}) as
\bea
\label{eq:G1Nquasiinbellaforma}
\frac{\xi_0}{{\gamma^{\rm wire}} G_{1N}} 
= 
{\xi_N} 
- \lrb \frac{\Sigma^{\rm L}}{{\gamma^{\rm wire}}} + 
\frac{\Sigma^{\rm R}}{{\gamma^{\rm
wire}}} \rrb
{\xi_{N-1}} 
+ \frac{\Sigma^{\rm L} \Sigma^{\rm R} }{{\gamma^{\rm wire}}^2} 
{\xi_{N-2}} .
\eea
In other words, the inverse Green function matrix element connecting left and right leads 
can be written as a sum of terms, representing the inverse of the bare
Green function matrix elements for a wire of $N$, $N-1$, and $N-2$ atoms.
In the limit of weak contact coupling, the behavior of the $G_{ 1 N}$ element 
is dominated by $\xi_{N}$, leading to $N$ transmission resonances in the 
conductance of unit height. Nevertheless, if the effective coupling between 
the molecule and the lead is much larger than $\gamma^{\rm wire}$, $\xi_{N-2}$ will 
become the dominant term. As a consequence the conductance spectrum is 
effectively that of an $(N\!-\!2)$-atomic wire~\cite{FCR01a}. 
The resonant behavior inside the wire band ($|\mcalE^{\rm wire} | \le 1$) and
its modification due to the lead coupling 
is easily understood by writing the
transmission as $T=4 \delta^2 \sin^2 (\vt) / \mcal{D}$ where the denominator $ \mcal{D}$ 
takes the following compact exact form valid for all $N \ge 1$:
\beann
\mcal{D}&=&
\lrb 
\sin(N+1)\vt 
-\lrb {\delta^2 - \lambda^2}\rrb
\sin(N-1)\vt 
-{2\lambda}
\sin N\vt 
\rrb^2
\\ && 
+ \ 4 \ \lrb 
\delta 
\sin N\vt 
-{\lambda \delta}
\sin(N-1)\vt 
\rrb^2 \, .
\eeann
Here $\sigma= \lambda - \ii \delta = \Sigma / \gamma^{\rm wire}$ is the self-energy 
of the leads (for simplicity assumed to be equal) normalized by the wire hopping. 
The parameter 
\bea
\label{eq:parametrization}
\vt = \cos^{-1} \mcalE^{\rm wire} = \frac{\ii}{2} \ln \frac {\mcalE^{\rm wire} - \sqrt{{\mcalE^{\rm wire}}^2 -1}} {\mcalE^{\rm wire} + \sqrt{{\mcalE^{\rm wire}}^2 -1}},
\eea
is real in the wire band giving rise to resonances for injected electrons
matching the wire eigenenergies. Outside the wire band $\vt$ is pure imaginary
($\sin$ functions are effectively sinh functions), and the transmission has a power law
dependence on energy and an exponential dependence on the wire length, that is
\begin{equation}
T \sim | 2 \mcalE^{\rm wire} |^{-2 N} \quad \mbox{for} \quad | \mcalE^{\rm
wire} | \gg1 \, ,
\end{equation}
in agreement with previous results~\cite{McConnell61}.
This analytic expression for the transmission provides the generalization 
of existing results~\cite{MKR94a+MKR94b,HRHS00,Sumetskii91+MI80} to the case with
non-vanishing real part of the self-energies.
The density of states $\mcal{N} = - {\rm Im~Tr}\lcb G \rcb / \pi $ can also be 
written in a closed analytical form. One can therefore take advantage of the fact 
that, due to the wire homogeneity, all the diagonal elements except the first 
and the last one coincide,
\beann
\left. \phantom{\frac {\frac 1 2 } {\frac 1 2 }} G_{kk} \; 
\rabs_{1<k<N}=\frac 1 {\gamma^{\rm wire}} \; \frac
{\xi_{N-1} -2 \sigma \xi_{N-2} + \sigma^2 \xi_{N-3}}
{\xi_N -2 \sigma \xi_{N-1} + \sigma^2 \xi_{N-2}} \; .
\eeann
By using the parametrization~(\ref{eq:parametrization}) one can easily
recast the density of states into the compact form
\beann
\mcal{N} &=& - \frac1 {\pi \gamma^{\rm wire}} \; \times
\\ & & 
{\rm Im} \; \frac
{N \sin N \vt -2 (N-1) \sigma \sin (N-1)\vt + (N-2) \sigma^2 \sin(N-2)\vt}
{\sin (N+1)\vt -2 \sigma \sin N \vt + \sigma^2 \sin(N-1)\vt} .
\eeann

\section{Electrode self-energy}
In calculating the spectral function, we 
make use of the assumption of identical left and right leads and drop the
self-energy indices in Eq.~(\ref{eq:spectr-density}).
Since the Hamiltonian is discrete, we can write the lattice 
Green function $G = \lrb E + {\rm i} 0^{+} - H \rrb^{-1}$
in matrix form by rearranging the two dimensional $n$ lattice coordinate in
Eq.~(\ref{eq:hamiltonian}).
We assume the $x$ direction to be parallel to the
tubes (and to the transport direction) and $y$ to be the finite transverse coordinate
(see Fig.~1).
The latter is curvilinear with $n_y$ spanning $M$ sites with periodic boundary
conditions.

The lattice representation of the lead Green function is needed in the calculation of
the self-energy contribution. 
It can generally be written by projecting the Green operator onto the localized
state basis,
$\psi_{k_x,k_y} (n_x = {\rm border}, n_y) = \chi_{k_x} \phi_{k_y} (n_y)$,
of the semi-infinite lead:
\bea
\mathcal{G}_{n_y^{\phantom{\prime}} n_y^\prime } \lrb E \rrb &=& \lab
n_y^{\phantom{\prime}} \rabs 
\lrb {E + \ii 0^+ -H} \rrb ^{-1} 
\labs n_y^\prime \rab \nonumber \\
\label{eq:lead_gf_general_problem}
&=&\sum_{k_x, k_y} \frac
{ \chi_{k_x} \phi_{k_y} (n_y^{\phantom{\prime}}) \chi^*_{k_x} \phi^*_{k_y}
(n_y^\prime)}
{E + \ii 0^+ - E_{k_x, k_y}}
\eea

\subsection{One-dimensional electrodes: Newns model}

We first recall the particular case of linear chain electrodes
(onto which we will map our system due to the validity of a channel selection).
For such a simple model the dispersion 
relation as a function of the lattice on-site energies
$\varepsilon$, of the hopping terms $\gamma$, and of the lattice spacing $a$ is simply given by
$E = \varepsilon - 2 \gamma \cos k_x a$.
As a result the surface Green
function for the semi-infinite chain is obtained by inserting the wave
function at the lead origin $\chi_{{k}_{x}} = \sqrt{2/\pi} \sin k_x a$ in the
defining expression~(\ref{eq:lead_gf_general_problem}) and transforming the
only sum over momenta into an integral, due to the infinite system size.
Thus,
\beann
\mathcal{G} \lrb E \rrb 
= 
\frac{a} \pi \int_{-\pi / a}^{\pi / a} dk_x \; \frac {\sin^2 k_x a}{E+\ii 0^+ -
\varepsilon + 2 \gamma \cos k_x a}= 
\frac {\ee^{\ii k_x (E) a}} {-\gamma},
\eeann
where we solved the integral according to Refs.~\cite{Brodovitsky58} 
and we made use of the dispersion relation. 
The resulting spectral density, given by Eq.~(\ref{eq:spectr-density}), is the 
semi-elliptical local density of states (LDOS) as obtained by Newns in his theory 
of chemisorption~\cite{Newns69}:
\beann
\Delta^{\rm Newns} \lrb \mcalE \rrb= \frac{\Gamma_{\rm eff}^2}{\gamma} \sqrt {1 - 
\mcalE^2 } \; \; \Theta \lrb {1 - \labs 
\mcalE \rabs } \rrb .
\eeann 
$\Gamma_{\rm eff}$ is the strength of the single contact between the
molecule and the semi-infinite one-dimensional leads,
$\mcalE=(E-\varepsilon)/(2\gamma)$ is the band-normalized energy, and $\Theta$
the Heaviside function.
The real part of the self-energy, responsible for shifting the molecular
resonances, is simply proportional to $\cos k_x a$ and thus linear in energy.
Its full dependence on energy is given by Hilbert transforming $\Delta$,
and it reads 
\beann
{\rm Re} \; \Sigma = \frac{\Gamma_{\rm eff}^2}{\gamma}
\lrb 
\mcalE -
\sqrt {
\mcalE^2 
-1} \; 
\lrb 
\Theta \lrb \mcalE -1 \rrb 
- \Theta \lrb - \mcalE -1 \rrb 
\rrb
\rrb .
\eeann

\subsection{Square lattice tubular electrodes}
Square lattice leads are characterized by periodic boundary 
conditions perpendicularly to the lead direction.
Transverse momentum quantization leads to 
$k_y^j a=2 \pi j / M$ (with $0 \le j < M$).
The surface Green function for such a system can be written as
\bea
\mathcal{G}_{n_y^{\phantom{\prime}} n_y^\prime} \lrb E\rrb
&=&
\frac {a}{\pi M} \sum_{k_y} 
\int_{-\pi / a}^{\pi / a} dk_x \; \frac {\sin^2 \lrb k_x a \rrb
\; 
\; 
\varphi_j^{\phantom{*}} \lrb n_y^{\phantom{\prime}} \rrb 
\varphi_j^* \lrb n_y^\prime \rrb ,
}{E+\ii 0^+ -
\varepsilon + 2 \gamma \cos k_y a + 2 \gamma \cos k_x a} 
\nonumber
\\
\label{eq:general-inf-green-funct}
&=& \frac{1}{M} \sum_{j=0}^{M-1}
\varphi_j^{\phantom{*}} \lrb n_y^{\phantom{\prime}} \rrb
\tilde{G}^j \lrb E \rrb
\varphi_j^* \lrb n_y^\prime \rrb ,
\eea
where $\tilde{G}^j \lrb E \rrb =
- {\ee^{\ii k_x^j (E) a}}/{\gamma}$ 
has been obtained by solving an integral formally equivalent to the linear
chain case and using the dispersion relation
\bea
\label{eq:disp-rel}
E = \varepsilon - 2 \gamma
\lrb \cos k_y^j a\ + \cos k_x^j (E) a \rrb .
\eea
The transverse profile of the wave function is given by
$\varphi_j (n_y) = %
\exp (\ii k_y^j n_y a)$.
Note that the wave function is obtained by a further
normalization, namely $\phi = \varphi / (Ma)^{1/2}$.

The self-energy finally reads as a sum of weighted longitudinal wave function profiles
\beann
\Sigma = \frac{1}{M} \sum_{j=0}^{M-1} 
\tilde{G}^j ( E )
\eta_{{j} / {M}} \lsb \Gamma \rsb ,
\eeann
where the weight
\bea
\label{eq:eta}
\eta_{{j} / {M}} \lsb \Gamma \rsb = \labs
\sum_{m=0}^{M-1} \Gamma_m
\varphi_j (m)
\rabs^2 
\eea
is the contact-averaged transverse wave function. Depending on the contact geometry
one has to specify the distribution of the $\Gamma_m$ contacts to calculate the
weight $\eta$ and thus the self-energy.
Note that $\eta_{(\cdot)}$ is formally the square modulus of the Fourier
series of $\Gamma_{(\cdot)}$; thus the zero-mode transverse momentum state 
$j=0$ contributes to $\eta$ with the square of the mean contact strength.
Due to the geometry of the lead surface, it is reasonable to assume a uniform 
distribution of contacts between the molecular wire and the electrodes.
For contacts of equal strength 
$\Gamma_m= \Gamma_{\rm eff} / \sqrt{M_{\rm c}}$, 
active on $M_{\rm c} \leq M$ sites, 
we obtain a modulation for the contributing channels governed by
\beann
\eta_{j/M} \lrb \Gamma_{\rm eff}, M_{\rm c} \rrb =
\Gamma_{\rm eff}^2 M_{\rm c}
\frac {{\rm sinc}^2 \lrb \pi j M_{\rm c} / M \rrb}
{{\rm sinc}^2 \lrb \pi j / M \rrb} ,
\eeann
where ${\rm sinc} (x \ne 0 ) \equiv {\sin } (x) /x$, and ${\rm sinc} (x = 0 ) \equiv 1$.
One can decompose the spectral density into a sum over the spectral densities of each 
state $j$. Namely $\Delta = \Delta^{(0)} \sum_j w_j (E)$, 
with $\Delta^{(0)}=\Gamma_{\rm eff}^2 M_{\rm c} /( \gamma M)$.
The channel weights are obviously independent upon rotation of the interfacial
coupling position as $\Sigma$ itself is.

\begin{figure}[t]
\centerline{\epsfig{file=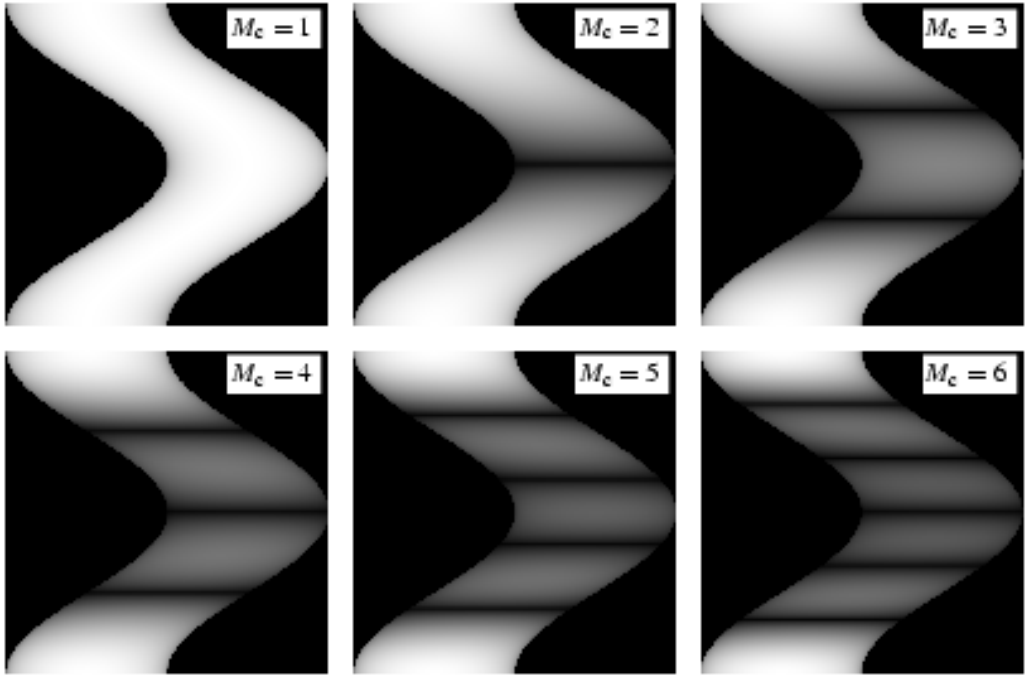, width=0.99\linewidth}}
\caption{\label{fig:fig1} Spectral weights $w_j (E)$ plotted for different 
contact values $M_{\rm c}$; a nonlinear grey level scale is used with black corresponding
to 0 and white to 1; small weights are amplified for better visualization. In every panel
the horizontal axis represents the normalized energy $-2 \le (E-\varepsilon)/(2 \gamma) 
\le 2$, and the vertical axis the normalized wave number $0 \le j/M < 1$. Note that in 
the mesoscopic limit, $M_c / M \lesssim 1$, the states $j \ne 0 $ match the 
nodes of $\Delta_j (E)$: only 
the zero-transverse momentum $j=0$ contributes to transport
(a better resolution figure is available upon request).}
\end{figure}

In Fig.~\ref{fig:fig1}., the weights $w_j (E)$ are visualized for different contact values $1 \le M_{\rm c} \le 6$.

For the case $M=M_{\rm c}$ the contributions from all states are suppressed
except the state with zero transverse momentum, which is the outcome of the sum 
rule (\ref{eq:eta}). That is, $\eta = \Gamma_{\rm eff}^2 M \delta_{j,0}$. 
Thus the configuration with all contacts of the tube ends coupled to the molecule 
with strength $\Gamma_{\rm eff} / \sqrt{M}$ is equivalent to the case of
a single contact with strength $\Gamma_{\rm eff}$ to a one-dimensional lead
(previous section). 
Moreover a scaling law is found for $\Sigma$, and {\it a fortiori} for the conductance 
given by $g = g \lrb \bar{\Gamma} \sqrt{M_{\rm c}} \rrb$, where
$\bar{\Gamma}$ is the local contact strength.

\begin{figure}[t]
\centerline{\epsfig{file=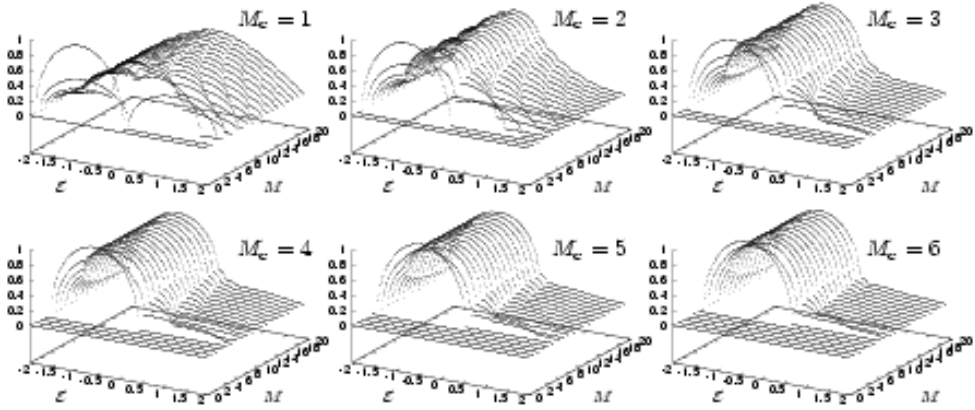, width=0.99\linewidth}}
\caption{\label{fig:fig2} The normalized spectral density $\Delta \gamma / \Gamma_{\rm eff}^2$ 
plotted for different maximum values $M_{\rm c}$ as in Fig.~\ref{fig:fig1}.; 
the $x$-axes represent the normalized 
energies $-2 \le (E-\varepsilon )/(2 \gamma) \le 2$, 
the $y$-axes label the dimensionality 
of the leads (number of possible contacts) $M_{\rm c} \le M \le 20$
(a better resolution figure is available upon request).}
\end{figure}

In Fig.~%
\ref{fig:fig2}%
, $\Delta$ is displayed as a function
of energy, lead diameters and active contacts. As easily visible, it is
only for values $M_{\rm c}$ of the order of the available contacts $M$ that
the mesoscopic nature of the scattering channels enter the spectral density. 
The larger the tube diameter the lower is the number of contacts which are
needed to reach a MC-like spectral density. 
This observation justifies the use of the one-dimensional Newns model for leads 
of lateral dimension much
larger than the contacted molecule but also shows the limit of this
approach when dealing with quasi-onedimensional leads. 
It remains to investigate to which extent the results obtained so far can
be generalized to realistic quasi-onedimensional structures such as CNT. 

\subsection{Carbon nanotube electrodes}
When the armchair ($\ell,\ell$) CNT topology is imposed the number of carbon sites at the interface is $M= 2 \ell$. 
The eigenvalues of the tight-binding hamiltonian (\ref{eq:hamiltonian}),
\bea
\label{eq:disp_rel_cnt}
E_{\pm} \lrb k_x^j,j\rrb = 
\varepsilon
\pm \gamma \sqrt {1 + 4\cos
\lrb \frac{ j
\phantom{k_x^j} \! \! \! \! \!
\pi}{\ell} \rrb \cos\lrb \frac{k_x^j a} 2 \rrb+ 4 \cos^2\lrb \frac{k_x^j a} 2 \rrb} \, ,
\eea
are obtained in a basis set given by symmetric ($+$) and antisymmetric ($-$) site
configurations of the graphene bipartite lattice, corresponding to 
$\pi$ and $\pi^*$ orbitals respectively~\cite{SFDD92,Wallace47}. The
longitudinal momentum is restricted to the Brillouin zone, $-\pi < k^j_x a < \pi$, and the transverse wave number $1 \le j \le 2 \ell$ labels $4 \ell$ bands, as many as the number of atoms in the unit cell of a $(\ell , \ell)$
CNT.
The two bands corresponding to $j=\ell$ are singly degenerate. They are responsible for 
the metallic character of armchair carbon nanotubes (these two bands cross at the 
Fermi level $E=\varepsilon$ for $k_x^\ell a = \pm 2 \pi /3$). Also the two outermost 
bands corresponding to $j=2 \ell$ are singly degenerate while the other remaining 
$(4 \ell -4)$ bands are collected in $(2 \ell -2)$ doubly-degenerate dispersion curves. 

\begin{figure}[t]
\centerline{\epsfig{file=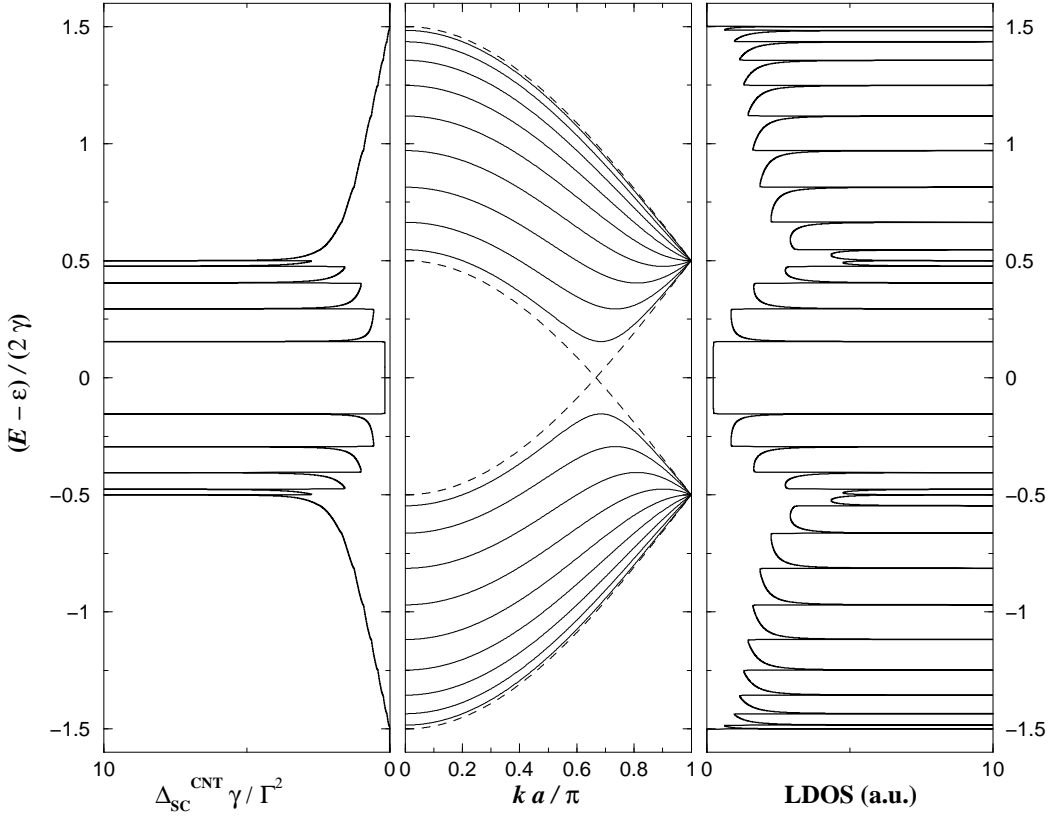, width=0.99\linewidth}}
\caption{\label{fig:ldos} 
Left panel: the normalized spectral density for a semi-infinite $(\ell, \ell)$ CNT lead in
the SC configuration; it corresponds to the LDOS at any atom site at the cut of the CNT 
lead. For comparison the dispersion relation and the
LDOS of an infinite $(\ell, \ell)$ CNT are shown in the middle and right
panel respectively. 
Solid lines in the dispersion relation panel indicate doubly degenerate bands,
dashed lines singly degenerate bands.
Here $\ell=10$, and on-site energies and hopping terms refer to $\alpha=\rm L,R$-leads.}
\end{figure}

\begin{figure}[t]
\centerline{\epsfig{file=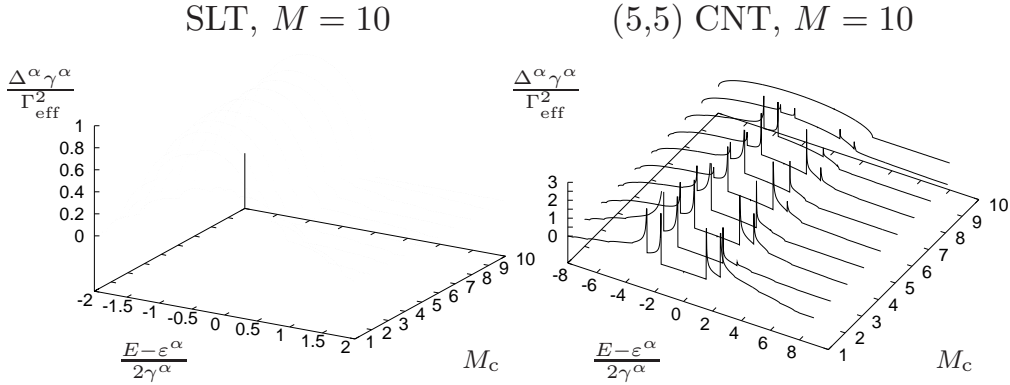, width=0.99\linewidth}}
\caption{\label{fig:spectrdens} The normalized spectral density as a function of energy and active contacts is plotted for $M=10$ possible atomic contacts available; on-site energies and hopping terms refer to $\alpha=\rm L,R$-leads.
The right panel illustrate numerical results after Ref.~\cite{CFR01a} in full agreement with the
analytics showed in the text.
}
\end{figure}

The single-particle Green function in a lattice representation 
for two sites belonging to the same sub-lattice
can be still written as in Eq.~(\ref{eq:general-inf-green-funct}) as
\bea
\label{eq:general-inf-green-funct-cnt}
\mathcal{G}_{n^{\phantom{\prime}}_y, n_y^\prime} \lrb E\rrb
= \frac{1}{2 \ell} \sum_{j=1}^{2\ell}
\varphi_j^{\phantom{*}} \lrb n^{\phantom{\prime}}_y \rrb
\tilde{G}^j \lrb 
E \rrb
\varphi_j^* \lrb n_y^\prime \rrb ,
\eea
where $\varphi_j (n_y) = %
\exp (\ii k_y^j n^{\phantom{j}}_y a)$, with $k_y^j a= \pi j / \ell$, and $1 \le j \le 2 \ell$.
Note that in 
Eq.~(\ref{eq:general-inf-green-funct-cnt}), $n_y^{\phantom{\prime}}$ {\it and}
$n_y^\prime$ should be either even or odd (that is they should belong to the
same sublattice).
The semi-infinite longitudinal Green function is given by 
\beann
\label{eq:lead_cnt_gf}
\tilde{G}^j \lrb E \rrb = 
\frac{a}{8 \pi}
\sum_{\beta=\pm}
\int_{-\pi/a}^{\pi/a}
dk_x^j \; 
\frac 
{\sin^2 \lrb k_x^j a / 2 \rrb}
{E + \ii 0^+ - E_\beta \lrb k_x^j,j\rrb} 
.
\eeann
The integral can be worked out analytically by extending $k_x^j$ to the complex plane 
and adding cross-cancelling paths (parallel to the imaginary axis) along the semi-infinite 
rectangle in the half plane ${\rm Im} k_x^j > 0$ and based on the interval between 
$-\pi/a$ and $\pi/a$. 
The closing path parallel to the real axis gives a real contribution linear in
energy.
This generalizes the approach by Ferreira {\it et al.}~\cite{FDML01},
recently adopted for obtaining an analytical expression for the diagonal Green function 
of infinite achiral tubes, to the case of semi-infinite CNTs.
The determination of the poles inside the integration contour, given by 
\beann
\cos \lrb \frac{q_\beta^j a}{2} \rrb = 
- \frac 1 2 
\cos \lrb \frac {j \pi}{\ell} \rrb 
- \frac{\beta} {2} \sqrt{\lrb \frac{E-\varepsilon}{2 \gamma}\rrb^2 - \sin^2
\lrb \frac{j \pi}{\ell} \rrb}
\eeann
allows for the calculation of the residues and thus of the surface Green function. 
One finds
\bea
\label{eq:surf_gf_cnt}
\tilde{G}^j \lrb E \rrb = 
\frac {1}{2 \gamma} \frac {E -\varepsilon} {2\gamma} 
\lrb 
 1 + \ii 
\frac{\ds{ 
\sin \lrb \frac{q_{\beta_*}^j a}{2} \rrb}}{\ds{ 
\sqrt{\lrb \frac{E-\varepsilon}{2 \gamma}\rrb^2 - \sin^2
\lrb \frac{j \pi}{\ell} \rrb}
}}
\rrb
\; ,
\eea
where the choice of the contributing pole through the branch parameter 
$\beta_* = {\rm sign} \, ( E -\varepsilon )$ has to be
taken into account.
The LDOS, obtained from the imaginary part of the surface Green
function after Eq.~(\ref{eq:surf_gf_cnt}) is plugged into
Eq.~(\ref{eq:general-inf-green-funct-cnt}), is shown in Fig.~\ref{fig:ldos}.
It clearly differs from the LDOS of an infinite CNT as depicted
for comparison in the right panel. As for the case of the SLT the pinning of the 
longitudinal wave function at the surface of the semi-infinite systems 
{\it cancels} all
border zone anomalies when $q_\pm^j a$ matches multiples of $2 \pi$.
In infinite SLTs these states are the {\it only} resonant states (van Hove
singularities) so
that the surface LDOS of a semi-infinite SLT never diverges (as it is
shown in the left panel of Fig.~\ref{fig:spectrdens}). On the contrary, in CNTs 
there are states with zero group velocity outside the border zone which
are responsible for the singularities of the spectral density of
semi-infinite CNTs (left panel of Fig.~\ref{fig:ldos}).

The self-energy for a CNT lead is more complicate than the one for a SLT 
owing to the missing equivalence of the sites belonging to the two different sub-lattices.
However, since the longitudinal part of the Green function,
Eq.~(\ref{eq:surf_gf_cnt}), is the same for all diagonal and off-diagonal terms
of the surface Green function,
the self-energy can still be cast into the form
\beann
\Sigma = \frac{1}{2 \ell} \sum_{j=1}^{2 \ell} 
\tilde{G}^j ( E )
\eta_{{j} / {\ell}} \lsb \Gamma \rsb .
\eeann
However, for the calculation of 
\bea
\label{eq:eta_cnt}
\eta_{{j} / {\ell}} \lsb \Gamma \rsb = \labs
\sum_{m=1}^{2 \ell} \Gamma_m
\varphi_j (m)
\rabs^2 ,
\eea
one has to specify the sub-lattice components of the transverse wave function
and whether they belong to a bonding or anti-bonding molecular state.
Again the distribution of the $\Gamma_m$ contacts is needed in oder to calculate the
weight $\eta$ and thus the self-energy.
Eq.~(\ref{eq:eta_cnt}) simplifies considerably in the SC case:
$\eta = \Gamma^2$. Since $\eta$ is uniform in $j$ 
the self-energy is simply proportional to the diagonal semi-infinite Green
function and, as a consequence, the spectral density is proportional to the 
local density of states~(Fig.~\ref{fig:ldos}.). 
The MC case ($\Gamma_m=\Gamma_{\rm eff}/\sqrt{2 \ell}$) is also easily tractable 
leading to a sum rule over the possible conducting channels.
However a direct proof is provided by the intuitive consideration that only
the $\pi$-bonding state can contribute to the MC spectral density 
(all the other states have a non-constant spatial modulation provided \eg \ in
Ref.~\cite{CI99}).
Following our notation the $\pi$-bonding state corresponds to $j=\ell$.
Fig.~(\ref{fig:spectrdens}) shows the spectral density 
in the intermediate regime between the SC and MC limits. The two
different lead lattice structures carry the same physical information only in
the MC limit case. 

\section{Discussion and concluding remarks}

It is interesting to recollect the results for the MC-spectral density, $\Delta_{\rm MC}$, in all
three lead models considered so far.
Using the dispersion relation~(\ref{eq:disp-rel}) and the surface Green function in 
Eq.~(\ref{eq:general-inf-green-funct}) 
the spectral density for SLT leads coincides formally with the Newns LDOS with an energy shift
$\Delta^{\rm SLT}_{\rm MC} (\mcalE)= \Delta^{\rm Newns} \lrb \mcalE + 1 \rrb$.
For CNTs Eqs.~(\ref{eq:disp_rel_cnt}-\ref{eq:surf_gf_cnt}) lead to 
$\Delta^{\rm CNT}_{\rm MC} (\mcalE)= \Delta^{\rm Newns} \lrb \mcalE + \frac 1 2 \rrb$.
From the above discussion it is clear that the multiple contact configuration suppresses features associated with the two-dimensional character of tubular leads,
apart from an energy shift.
In our model, the latter is the only remnant the system preserves 
from the transverse momentum component.

In contrast, the SC case is strongly dependent on the lead underlying
structure. The spectral density for a single contact, $\Delta_{\rm SC}$,
reduces to the LDOS in the lead at the point where the molecular wire 
is contacted with strength $\Gamma$,
\beann
\Delta_{\rm SC} = \pi \Gamma^2 \; \; {\rm LDOS}.
\eeann
It is, in particular, in the SC scenario that the conductance of the molecular
wire might be strongly affected by the nature of the leads~\cite{FCR01a}. Nevertheless, once
the nature of the contact can be inferred, one can think to cure the
spurious insertion in the conductance by filtering out the contribution of the 
leads from the molecular resonances. For instance,
in the CNT-enhanced STM tips~\cite{WMSS01} the improved resolution images can be cleaned 
by de-convolving them using model assumptions for the leads and their contact geometry.

Another significant consequence of the peculiar contact dependence of the
spectral density is the possibility
to understand the influence of the mesoscopic character of the leads.
In the limit of large $M$ (at fixed $M_{\rm c}$), $\eta$ looses
its granularity being sampled by many more states compared to its nodes,
whereas for $M_{\rm c} / M \lesssim 1$ an increasing number of nodes matches
the decreasing number of states.
This determines a reduction in the self-energy, and thus in the width of 
the molecular resonances, highlighting the quantum features of the wire.
The latter result in a quite striking behavior for CNTs because of the band
anomalies outside of the border zone which strongly determine the resonant
behavior of the spectral density.

To conclude, we have shown that
novel features are expected to arise in the conductance of a molecular wire
connected to nanotube leads. The commonly used approximation of a pure imaginary, 
flat, wide band self-energy is not valid when employing tubular leads.
Nevertheless, the conductance of a homogeneous molecular wire still possesses an 
analytical form in the entire regime of the wire parameters and allows for the 
insertions of a nonvanishing real self-energy, necessarily arising when considering 
nanotube leads. 
By tailoring the geometry and dimensionality of the contacts, it is possible to 
perform a channel selection. In the MC limit the conductance becomes independent 
of the lattice structure of the tubular electrodes, transport is dominated by 
topology properties and is effectively one-dimensional. 
Furthermore, the conductance obeys a universal scaling law in the multiple contact 
configuration.
We would further like to stress that the derived analytical expression for the semi-infinite 
CNT self-energy allows for a full analytical treatment of the linear conductance problem. 
The possibility to handle an exact expression of the semi-infinite CNT Green function 
may serve as a first step in analytical treatments of more complex carbon based 
molecular structures such as T- or Y-junctions~\cite{PZXWGP00+PRLVX00}.
\section{Acknowledgments}
Fruitful discussions and valuable correspondence with M.~S.~Ferreira are gratefully 
acknowledged.
R.~Guti{\'e}rrez and H.-S.~Sim provided perceptive comments to this manuscript. 
GC research at MPI is sponsored by the Schloe{\ss}mann Foundation.
GF acknowledges support from the Alexander von Humboldt Foundation.


{\small

\begin{thebibliography}{10}
\expandafter\ifx\csname url\endcsname\relax
\def\url#1{\texttt{#1}}\fi
\expandafter\ifx\csname urlprefix\endcsname\relax\def\urlprefix{URL }\fi

\bibitem{Taur97}
Y.~Taur, Proc.~IEEE 85 (1997) 4.

\bibitem{CDLFBLSKL00+SKBB00+Ziemelis98+Tour96}
B.~Crone, A.~Dodabalapur, Y.~Lin, R.~W.~Filas, Z.~Bao, A.~Laduca,
R.~Sarpeshkar, H.~E.~Katz, W.~Li, Nature 403 (2000) 521-523;
%
J.~H.~Sch{\"o}n, C.~Kloc, E.~Bucher, B.~Batlogg, Nature 403 (2000) 408-410;
%
K.~Ziemelis, Nature 394 (1998) 619-620;
%
J.~M.~Tour, Chem.~Rev. 96 (1996) 537-554.

\bibitem{EL00}
J.~C.~Ellenbogen, J.~C.~Love, Proc.~IEEE 88 (2000) 386-426.

\bibitem{TKS98}
J.~M.~Tour, M.~Kozaki, J.~M.~Seminario, J.~Am.~Chem.~Soc. 120 (1998)
8486-8493.

\bibitem{G-GMLOL97}
D.~{Goldhaber-Gordon}, M.~S.~Montemerlo, J.~C.~Love, G.~J.~Opiteck, J.~C.
Ellenbogen, Proc.~IEEE 85 (1997) 521-540.

\bibitem{Landauer96c}
R.~Landauer, IEEE Trans.~Electron Devices 43 (1996) 1637-1639.

\bibitem{AR74a}
A.~Aviram, M.~A.~Ratner, Chem.~Phy.~Lett. 29 (1974) 277-283.

\bibitem{WB93+MSA93}
D.~H.~Waldeck, D.~N.~Beratan, Science 261 (1993) 576-577;
%
A.~S.~Martin, J.~R.~Sambles, G.~J.~Ashwell, Phys.~Rev.~Lett. 70 (1993)
218-221.

\bibitem{RZMBT97}
M.~A.~Reed, C.~Zhou, C.~J.~Muller, T.~P.~Burgin, J.~M.~Tour, Science 278 (1997)
252-254.

\bibitem{PBdeVD00}
D.~Porath, A.~Bezryadin, S.~de~Vries, C.~Dekker, Nature 403 (2000) 635-638.

\bibitem{Nitzan01a}
A.~Nitzan, Ann.~Rev.~Phys.~Chem. 52 (2001) 681-750.

\bibitem{ROBWML01}
J.~Reichert, R.~Ochs, D.~Beckmann, H.~B.~Weber, M.~Mayor, and H.~v.~L\"ohneysen,
To appear in Phys. Rev. Lett.; \texttt{cond-mat/0106219}.

\bibitem{vanRuitenbeek01}
J.~M. {van Ruitenbeek}, Naturwissenschaften 88 (2001) 59-66.

\bibitem{YRGvdBAvR98}
A.~I.~Yanson, G.~{Rubio Bollinger}, H.~E. {van den Brom}, N.~Agra{\"\i}t, J.~M. {van
Ruitenbeek}, Nature 395 (1998) 783-785.

\bibitem{OKT98}
H.~Ohnishi, Y.~Kondo, K.~Takayanagi, Nature 395 (1998) 780-783.

\bibitem{KKKCSRO01}
A.~Karlsson, R.~Karlsson, M.~Karlsson, A.~Cans, A.~Str{\"o}mberg,
F.~Rytts{\'e}n, O.~Orwar, Nature 409 (2001) 150-152.

\bibitem{RKJTCL00+YWTA01}
T.~Rueckes, K.~Kim, E.~Joselevich, G.~Y.~Tseng, C.-L.~Cheung, C.~M.~Lieber,
Science 289 (2000) 94-97;
N.~Yoneya, E.~Watanabe, K.~Tsukagoshi, Y.~Aoyagi, App.~Phys.~Lett. 79
(2001) 1465.

\bibitem{SDD98+McEuen00}
R.~Saito, G.~Dresselhaus, M.~S.~Dresselhaus, Physical Properties of Carbon
Nanotubes, World Scientific Publishing, London, 1998;
%
P.~{McEuen}, Phys.~World 13 (2000) 31-36.

\bibitem{TMDSRPNB01}
C.~Thelander, M.~H.~Magnusson, K.~Deppert, L.~Samuelson, P.~{Rugaard Poulsen},
J.~Nyg{\aa}rd, J.~Borggreen, App.~Phys.~Lett. 79 (2001) 2106-2108.

\bibitem{ADX00+dePGWADR99}
J.~Hu, M.~Ouyang, P.~Yang, C.~M.~Lieber, Nature 399 (1999) 48-51;
P.~J. {de Pablo}, E.~Graugnard, B.~Walsh, R.~P.~Andres, S.~Datta,
R.~Reifenberger, App.~Phys.~Lett. 74 (1999) 323-325.

\bibitem{DMAA01+PTYGD01+MSSHA98}
V.~Derycke, R.~Martel, J.~Appenzeller, Ph.~Avouris, Nano Letters 1 (2001) 453-456;
%
H.~W.~C.~Postma, T.~Teepen, Z.~Yao, M.~Grifoni, C.~Dekker, Science 293 (2001)
76-79;
%
R.~Martel, T.~Schmidt, H.~R.~Shea, T.~Hertel, Ph.~Avouris, App.~Phys.~Lett.
73 (1998) 2447-2449.

\bibitem{WMSS01}
H.~Watanabe, C.~Manabe, T.~Shigematsu, M.~Shimizu, App.~Phys.~Lett. 78 
(2001) 2928-2930.

\bibitem{WJWCL98}
S.~S.~Wong, E.~Joselevich, A.~T.~Woolley, C.~L.~Cheung, C.~M.~Lieber, Nature
394 (1998) 52-55.

\bibitem{NKANHYT00}
H.~Nishijima, S.~Kamo, S.~Akita, Y.~Nakayama, K.~I.~Hohmura, S.~H.~Yoshimura,
K.~Takeyasu, App.~Phys.~Lett. 74 (2000) 4061-4063.

\bibitem{ANKN01+OBKMRGvdDR00+VdePHMLYMMRF98}
S.~Akita, H.~Nishijima, T.~Kishida, Y.~Nakayama, Jpn.~J.~Appl.~Phys. 39
(2000) 7086-7089;
%
A.~I.~Onipko, K.-F.~Berggren, Y.~O.~Klymenko, L.~I.~Malysheva, J.~J.~W.~M.
Rosink, L.~J.~Geerligs, E.~{van der Drift}, S.~Radelaar, Phys.~Rev.~B 61 
(2000) 11118-11124;
%
A.~L. {V{\'a}zquez de Parga}, O.~S.~Hern{\'a}n, R.~Miranda, A.~{Levy Yeyati},
N.~Mingo, A.~{Mart{\'\i}n-Rodero}, F.~Flores, Phys.~Rev.~Lett. 80 (1998)
357-360.

\bibitem{CGFGRS01}
G.~Cuniberti,~R.~Guti{\'e}rrez, G.~Fagas, F.~Gro{\ss}mann, K. ~Richter, and R.~Schmidt,
Physica E 12 (2002) 749;
R.~Guti{\'e}rrez, G.~Fagas, G.~Cuniberti, F.~Gro{\ss}mann, R.~Schmidt, and K.~Richter,
Phys. Rev. B 65 (2002) 113410.

\bibitem{LA00}
N.~D.~Lang, Ph.~Avouris, Phys.~Rev.~Lett. 84 (2000) 358-361.

\bibitem{SFDD92}
R.~Saito, M.~Fujita, G.~Dresselhaus, M.~S.~Dresselhaus, Phys.~Rev.~B 46 
(1992) 1804-1811.

\bibitem{IL99}
Y.~Imry, R.~Landauer, Rev.~Mod.~Phys. 71 (1999) S306.

\bibitem{FG99}
D.~K.~Ferry, S.~M.~Goodnick, Transport in Nanostructures, Vol.~6 of Cambridge
Studies in Semiconductor Physics \& Microelectronic Engineering, Cambridge
University Press, Cambridge, 1999.

\bibitem{FL81}
D.~S.~Fisher, P.~A.~Lee, Phys.~Rev.~B 23 (1981) R6851-R6854.

\bibitem{MKR94a+MKR94b}
V.~Mujica, M.~Kemp, M.~A.~Ratner, J.~Chem.~Phys. 101 (1994) 6849-6855;
%
{\it ibid.} pag. 6856-6864.
%

\bibitem{HRHS00}
L.~E.~Hall, J.~R.~Reimers, N.~S.~Hush, K.~Silverbrook, J.~Chem.~Phys. 112 
(2000) 1510-1521.

\bibitem{SO93}
S.~S.~Skourtis, J.~N.~Onuchic, Chem.~Phy.~Lett. 209 (1993) 171-177.

\bibitem{FCR01a}
G.~Fagas, G.~Cuniberti, K.~Richter, Phys.~Rev.~B 63 (2001) 045416.

\bibitem{McConnell61}
H.~M.~McConnell, J.~Chem.~Phys. 35 (1961) 508-515.

\bibitem{Sumetskii91+MI80}
M.~Sumetskii, J.~Phys.-Condens.~Matter 3 (1991) 2651-2654;
%
V.~V.~Malov, L.~V.~Iogansen, Opt.~Spektrosk. (USSR) 48 (1980) 146-154.

\bibitem{Brodovitsky58}
K.~V.~Brodovitsky, Doklady Akademii NAUK SSSR 120 (1958) 1178-1179; see also 
eq.~3.644.4 in I.~S.~Gradschteyn, I.~M.~Ryzhik, Tables of Integrals, Series,
and Products, 6th Edition, Academic Press, Inc., San Diego, 2000.


\bibitem{Newns69}
D.~M.~Newns, Phys.~Rev. 178 (1969) 1123.

\bibitem{Wallace47}
P.~R.~Wallace, Phys.~Rev. 71 (1947) 622-634.

\bibitem{FDML01}
M.~S.~Ferreira, T.~G.~Dargam, R.~B.~Muniz, A.~Latg{\'e}, Phys.~Rev.~B 63 (2001)
245111; this reference contains some misprints, especially in the final diagonal form of the Green function of an infinite CNT amended in the erratum
{\it ibid.} Phys.~Rev.~B 65 (2002) 039901.

\bibitem{CFR01a}
G.~Cuniberti, G.~Fagas, K.~Richter, Acta Phys.~Pol. 32 (2001) 437-442.

\bibitem{CI99}
H.~J.~Choi, J.~Ihm, Solid State Comm. 111 (1999) 385-390.

\bibitem{PZXWGP00+PRLVX00}
L.-M.~Peng, Z.~L.~Zhang, Z.~Q.~Xue, Q.~D.~Wu, Z.~N.~Gu, D.~G.~Pettifor, Phys.~Rev.~Lett. 85~(15) (2000) 3249-3252;
%
C.~Papadopoulos, A.~Rakitin, J.~Li, A.~S.~Vedeneev, J.~M.~Xu, Phys.~Rev.~Lett. 85~(16) (2000) 3476-3479.

\end{thebibliography}

}
\end{document}